\documentclass[journal]{IEEEtran}

\usepackage[T1]{fontenc}
\usepackage{amsmath,amssymb,amsthm}
\usepackage{cite}
\usepackage{balance}
\usepackage[hidelinks]{hyperref}
\usepackage{microtype}
\hypersetup{pdftitle={Length Bounds for MDS Algebraic Geometry Codes of Arbitrary Positive Genus}}

\newtheorem{theorem}{Theorem}
\newtheorem{corollary}[theorem]{Corollary}
\newtheorem{proposition}[theorem]{Proposition}
\newtheorem{lemma}[theorem]{Lemma}
\theoremstyle{remark}
\newtheorem{remark}[theorem]{Remark}
\newcommand{\Fq}{\mathbb F_q}
\newcommand{\C}{\mathbb C}
\DeclareMathOperator{\supp}{supp}
\DeclareMathOperator{\Jac}{Jac}

\DeclareMathOperator{\res}{res}

\begin{document}

\title{Length Bounds for MDS Algebraic Geometry Codes\protect\\
of Arbitrary Positive Genus}
\author{Haojie Chen, Chuangqiang Hu, and Guorui Ma%
\thanks{Haojie Chen and Chuangqiang Hu are with Sun Yat-sen University, Guangzhou 510275, China (e-mail: chenhj69@mail2.sysu.edu.cn; huchq5@mail.sysu.edu.com).}%
\thanks{Guorui Ma is with the Shanghai Institute for Mathematics and Interdisciplinary Sciences (SIMIS), Shanghai 200433, China, and the Research Institute of Intelligent Complex Systems, Fudan University, Shanghai 200433, China (e-mail: mgr18@tsinghua.org.cn). Guorui Ma is the corresponding author.}}
\markboth{Length Bounds for MDS Algebraic Geometry Codes}{Length Bounds for MDS Algebraic Geometry Codes}
\maketitle

\begin{abstract}
We study the length of maximum distance separable algebraic geometry codes from curves of arbitrary positive genus. For fixed genus, we prove that the asymptotic ratio of the length to the field size is at most $1/2$ when both the dimension and the codimension grow faster than the logarithm of the field size. We also give explicit bounds under logarithmic dimension conditions. The bounds hold in every characteristic and are uniform in the curve and the evaluation set. The proof combines character sums on the Jacobian with an estimate for sums of distinct points.
\end{abstract}


\section{Introduction}
\label{sec:introduction}

\IEEEPARstart{A}{ linear} $[n,k,d]_q$ code is maximum distance separable (MDS) if it meets the Singleton bound $d\leq n-k+1$ with equality \cite{Singleton1964}. Determining the possible lengths of MDS codes is a basic problem in coding theory. For $2\leq k\leq q-1$, the classical linear MDS conjecture predicts $n\leq q+1$. The exceptions allow $n=q+2$ when $q$ is even and $k\in\{3,q-1\}$; see \cite[Ch.~11]{MacWilliamsSloane1977}. For codes defined by algebraic curves, the geometry of the curve gives further constraints on the length.

Algebraic geometry codes are obtained by evaluating functions in a Riemann--Roch space at distinct rational points. On the projective line, this construction gives generalized Reed--Solomon codes. On a curve of positive genus, the same construction can produce MDS codes, but the defining divisor must satisfy additional vanishing conditions \cite[Ch.~2]{Stichtenoth2009}. These conditions relate the length problem to the arithmetic of divisor classes.

Munuera \cite{Munuera1992} used this relation to study the MDS conjecture for geometric codes, including codes from curves of genera one and two. Chen and Yau \cite{ChenYau1997} developed the method for hyperelliptic curves and obtained the $q+1$ bound under explicit conditions on the field size. These results suggest a more precise question: for a fixed positive genus, how far below $q+1$ must the length lie?

For elliptic curves, the group law gives a direct relation between the MDS property and sums of distinct rational points. Li, Wan, and Zhang \cite{LiWanZhang2015} proved a length bound with asymptotic coefficient $2/3$ when both the dimension and the codimension are larger than a suitable multiple of $\log q$. They conjectured that the coefficient could be reduced to $1/2$. Han and Ren \cite{HanRen2024} proved the corresponding asymptotic bound for nontrivial MDS elliptic codes. Further exact results in restricted dimension ranges were obtained by Chen, Hu, Huang, and Zhao \cite{ChenHuHuangZhao2026}.

For a curve of genus $g>1$, its rational points do not form a group. They do, however, map to the group of rational points of its Jacobian. For fixed $g$, this group has order about $q^g$, whereas the curve has about $q$ rational points. Thus an evaluation set is usually a small subset of the Jacobian. A bound for sums of distinct points must use the fact that these points lie on a curve. Character sums provide this information. Peng, Chen, and Zhao \cite{PengChenZhao2022} used this approach in algebraic geometric secret sharing. Their result as $q\to\infty$ assumes that only a bounded number of rational points are omitted from the evaluation set. Here we allow arbitrary evaluation sets. We use a coefficient estimate to obtain a condition under which every divisor class of a prescribed degree has a representative supported on distinct evaluation points. This condition gives the length bound near $q/2$.

The aim of this paper is to obtain a bound near $q/2$ for curves of arbitrary positive genus. We allow the dimension and the codimension to grow with $q$. This permits a proof that does not use a map of small degree to the projective line. The geometric input consists of the Weil bounds for the curve, its Jacobian, and nontrivial characters on the curve. We combine these bounds with a coefficient estimate that counts subsets without repeated points.

We now state the main result. Let $X/\Fq$ be a smooth projective geometrically irreducible curve of genus $g\geq1$. Let $D=P_1+\cdots+P_n$, where the $P_i$ are distinct rational points, and let $G$ be an $\Fq$-rational divisor with support disjoint from $D$. We write $C_{\mathcal L}(D,G)$ for the evaluation code defined by the full space $\mathcal L(G)$, and put
\begin{equation}
    k=\dim_{\Fq}C_{\mathcal L}(D,G),
    \qquad \kappa=\min\{k,n-k\}.
    \label{eq:kappa}
\end{equation}
All logarithms in this paper are natural logarithms. Define
\begin{equation}
\begin{split}
    \Lambda_g(q)={}&2g\log(\sqrt q+1)\\
                 &+\log(q+2+2g\sqrt q).
\end{split}
\label{eq:Lambda}
\end{equation}

\begin{theorem}
\label{thm:main}
Suppose that $C_{\mathcal L}(D,G)$ is MDS and that
\begin{equation}
    \kappa\geq g+1,\qquad \kappa>\Lambda_g(q).
    \label{eq:main-hypotheses}
\end{equation}
Then
\begin{equation}
    n\leq
    \frac{q+1+(4g-2)\sqrt q}
         {2\bigl(1-\Lambda_g(q)/\kappa\bigr)}.
    \label{eq:main-bound}
\end{equation}
\end{theorem}

The theorem is uniform in the curve, the evaluation set, and the divisor. It holds in every characteristic. No degree condition on $G$ is imposed in the statement. The condition $\kappa\geq g+1$ implies $2g\leq\deg G=k+g-1<n$, as proved in Section~\ref{sec:preliminaries}. The support of $G$ may contain closed points of degree greater than one.

The next consequence states the bound with a prescribed coefficient greater than $1/2$. It also gives the required dimension range explicitly.

\begin{corollary}
\label{cor:epsilon}
Let $0<\varepsilon<1/2$, and set
\begin{equation}
    K_{g,\varepsilon}(q)=
    \max\left\{g+1,
    \left\lceil\frac{1+2\varepsilon}{2\varepsilon}
                  \Lambda_g(q)\right\rceil\right\}.
    \label{eq:epsilon-threshold}
\end{equation}
If $C_{\mathcal L}(D,G)$ is MDS and
\begin{equation}
    K_{g,\varepsilon}(q)\leq k\leq n-K_{g,\varepsilon}(q),
    \label{eq:epsilon-range}
\end{equation}
then
\begin{equation}
    n\leq\left(\frac12+\varepsilon\right)
            \bigl(q+1+(4g-2)\sqrt q\bigr).
    \label{eq:epsilon-bound}
\end{equation}
For fixed $g$ and $\varepsilon$,
\begin{equation}
    K_{g,\varepsilon}(q)
    =\frac{1+2\varepsilon}{2\varepsilon}(g+1)\log q
       +O_{g,\varepsilon}(1).
    \label{eq:threshold-asymptotic}
\end{equation}
\end{corollary}

When the dimension and the codimension are both a fixed positive fraction of the length, the coefficient $1/2$ can be retained without an $\varepsilon$ loss.

\begin{corollary}
\label{cor:rate}
Let $0<\eta\leq1/2$. Suppose that $C_{\mathcal L}(D,G)$ is MDS and
\begin{equation}
    \kappa\geq g+1,\qquad \kappa\geq\eta n.
    \label{eq:rate-hypotheses}
\end{equation}
Then
\begin{equation}
    n\leq\frac{q+1}{2}+(2g-1)\sqrt q
                    +\frac{\Lambda_g(q)}{\eta}.
    \label{eq:rate-bound}
\end{equation}
\end{corollary}

For fixed genus, the last term in \eqref{eq:rate-bound} has logarithmic order. We state this asymptotic form separately.

\begin{corollary}
\label{cor:fixed-genus}
Fix an integer $g\geq1$ and a real number $0<\eta\leq1/2$. Let $C_{\mathcal L}(D,G)$ be an MDS code from a curve of genus $g$ over $\Fq$. If $\kappa\geq g+1$ and $\kappa\geq\eta n$, then, as $q\to\infty$,
\begin{equation}
    n\leq\frac q2+(2g-1)\sqrt q+O_{g,\eta}(\log q).
    \label{eq:rate-asymptotic}
\end{equation}
The implied constant depends only on $g$ and $\eta$.
\end{corollary}

These statements concern complete evaluation codes in the displayed dimension ranges. They do not give a bound near $q/2$ for every fixed value of $k$ or $n-k$. In genus one, the results of Han and Ren \cite{HanRen2024} cover a wider range of dimensions. Our purpose is a bound with explicit dependence on the genus that applies to every curve of positive genus.

The paper is organized as follows. Section~\ref{sec:preliminaries} recalls the geometric and coding facts used in the proof. Section~\ref{sec:subset-sums} proves a subset sum estimate and a necessary condition for the MDS property. Section~\ref{sec:proof} proves the main theorem and its three corollaries. Section~\ref{sec:future} gives several questions for further work.

\section{Preliminaries}
\label{sec:preliminaries}

Throughout the paper, $X/\Fq$ is a smooth projective geometrically irreducible curve of genus $g\geq1$. Its function field is denoted by $\Fq(X)$. We first recall the facts needed to express the MDS condition in terms of divisors. We then state the character bound that will count the relevant divisors.

\subsection{Divisors and Riemann--Roch Spaces}

An $\Fq$-rational divisor is a finite sum
$A=\sum_P n_P P$ over the closed points of $X$, with $n_P\in\mathbb Z$. Its degree is $\deg A=\sum_P n_P\deg P$. It is effective, written $A\geq0$, if all $n_P$ are nonnegative. For $f\in\Fq(X)^\times$, let $(f)$ be its principal divisor. Two divisors are linearly equivalent if their difference is principal. The class of $A$ is denoted by $[A]$.

The Riemann--Roch space and its dimension are
\begin{equation}
\begin{split}
    \mathcal L(A)&=\{0\}\cup
       \{f\in\Fq(X)^\times:(f)+A\geq0\},\\
    \ell(A)&=\dim_{\Fq}\mathcal L(A).
\end{split}
\label{eq:RR-space}
\end{equation}
For a canonical divisor $K_X$, one has $\deg K_X=2g-2$. Riemann--Roch gives
\begin{equation}
    \ell(A)=\deg A+1-g+\ell(K_X-A).
    \label{eq:RR}
\end{equation}
Clifford's theorem states that, if $\ell(A)>0$ and $\ell(K_X-A)>0$, then
\begin{equation}
    \ell(A)\leq\frac{\deg A}{2}+1.
    \label{eq:Clifford}
\end{equation}
We use these results in their standard form; see \cite[Ch.~IV, Sec.~1]{Hartshorne1977}. They imply that
\begin{equation}
    \ell(A)\leq g\quad\text{if}\quad\deg A\leq2g-1.
    \label{eq:small-degree}
\end{equation}
Recall that $A$ is special if $\ell(K_X-A)>0$. In this case $\deg A\leq2g-2$, so \eqref{eq:small-degree} follows from \eqref{eq:Clifford} when $\ell(A)>0$. For a nonspecial divisor it follows from \eqref{eq:RR}. The assertion is immediate when $\ell(A)=0$.

\subsection{Evaluation Codes and the MDS Condition}

Let $P_1,\ldots,P_n\in X(\Fq)$ be distinct, put $D=P_1+\cdots+P_n$, and suppose that $\supp(G)\cap\supp(D)=\varnothing$. The evaluation code is
\begin{equation}
    C_{\mathcal L}(D,G)=
    \{(f(P_1),\ldots,f(P_n)):f\in\mathcal L(G)\}.
    \label{eq:evaluation-code}
\end{equation}
Its dimension is $\ell(G)-\ell(G-D)$. If $\deg G<n$, evaluation is injective and the dimension is $\ell(G)$.

We also use the duality formula for evaluation codes. A differential $\omega$ can be chosen so that
\[
    v_{P_i}(\omega)=-1,
    \qquad \res_{P_i}(\omega)=1
    \quad(1\leq i\leq n).
\]
Then
\begin{equation}
    C_{\mathcal L}(D,G)^\perp
       =C_{\mathcal L}(D,D-G+(\omega)).
    \label{eq:duality}
\end{equation}
The divisor on the right has support disjoint from $D$ and degree $n-\deg G+2g-2$. These facts follow from the standard duality theorem; see \cite[Sec.~2.2]{Stichtenoth2009}.

The next proposition explains the role of the lower bounds on the dimension and the codimension. It also gives the precise vanishing condition used later.

\begin{proposition}
\label{prop:code-divisors}
Suppose that $\dim C_{\mathcal L}(D,G)=k$ and that $k,n-k\geq g+1$. Then
\begin{equation}
    2g\leq\deg G=k+g-1<n,
    \qquad \ell(G)=k.
    \label{eq:degree-dimension}
\end{equation}
Under these assumptions, the code is MDS if and only if
\begin{equation}
    \ell(G-B)=0
    \quad\text{for every }0\leq B\leq D
    \text{ with }\deg B=k.
    \label{eq:mds-vanishing}
\end{equation}
\end{proposition}

\begin{proof}
If $\deg G\geq n$, the divisor in \eqref{eq:duality} has degree at most $2g-2$. By \eqref{eq:small-degree}, the dual code has dimension at most $g$. This contradicts $n-k\geq g+1$. Hence $\deg G<n$, and evaluation is injective. Thus $k=\ell(G)$. If $\deg G\leq2g-1$, \eqref{eq:small-degree} contradicts $k\geq g+1$. We obtain $\deg G\geq2g$, and \eqref{eq:RR} gives \eqref{eq:degree-dimension}.

A nonzero element of $\mathcal L(G-B)$ gives a nonzero codeword vanishing at the $k$ points in $B$. Its weight is at most $n-k$. Conversely, a nonzero codeword of weight at most $n-k$ vanishes at some $k$ evaluation points and gives such a divisor $B$. Therefore \eqref{eq:mds-vanishing} is equivalent to $d\geq n-k+1$. The Singleton bound gives equality.
\end{proof}

\subsection{Jacobians and Character Sums}

To find a divisor $B$ that violates \eqref{eq:mds-vanishing}, we will prescribe its class. Fix $P_0\in X(\Fq)$, and write
\begin{equation}
    J=\Jac(X)(\Fq),\qquad h=|J|,
    \qquad N=|X(\Fq)|.
    \label{eq:J-h-N}
\end{equation}
The group $J$ is identified with the degree-zero divisor classes over $\Fq$. The Abel--Jacobi map on rational points is
\begin{equation}
    \iota:X(\Fq)\longrightarrow J,
    \qquad P\longmapsto[P-P_0].
    \label{eq:Abel-Jacobi}
\end{equation}
This map is injective for $g\geq1$. Otherwise, distinct points $P,Q$ would satisfy $P\sim Q$, which would give a function of degree one and force $X$ to have genus zero.

The Weil bounds give
\begin{equation}
    |N-(q+1)|\leq2g\sqrt q,
    \qquad h\leq(\sqrt q+1)^{2g}.
    \label{eq:Weil}
\end{equation}
For the second inequality, the numerator of the zeta function is $\prod_{j=1}^{2g}(1-\alpha_j T)$, with $|\alpha_j|=\sqrt q$, and its value at $T=1$ is $h$. See \cite[Ch.~5]{Stichtenoth2009} and \cite[Ch.~9]{Rosen2002}.

A character of a finite abelian group $A$ is a homomorphism $\chi:A\to\C^\times$. Its values have absolute value one. We denote the character group by $\widehat A$ and its trivial character by $1$. The next standard estimate controls characters after restriction from the Jacobian to the curve.

\begin{theorem}
\label{thm:character-Weil}
For each nontrivial $\chi\in\widehat J$,
\begin{equation}
    \left|\sum_{P\in X(\Fq)}\chi([P-P_0])\right|
       \leq(2g-2)\sqrt q.
    \label{eq:curve-character}
\end{equation}
\end{theorem}

For $g\geq2$, this is the Weil bound for unramified character sums; see \cite[Ch.~9]{Rosen2002} and \cite[Prop.~2, preprint version]{PengChenZhao2022}. For $g=1$, the map \eqref{eq:Abel-Jacobi} identifies $X(\Fq)$ with $J$, and the sum is zero by character orthogonality.

For a subset $S\subseteq A$ of a finite abelian group with $|A|\geq2$, set
\begin{equation}
    \Phi(S)=\max_{\substack{\chi\in\widehat A\\\chi\neq1}}
                  \left|\sum_{a\in S}\chi(a)\right|.
    \label{eq:Phi}
\end{equation}
This standard quantity also occurs in subset sum estimates such as \cite{LiWanZhang2015}. For the image of our evaluation set,
\begin{equation}
    S=\{[P_i-P_0]:1\leq i\leq n\}\subseteq J,
    \label{eq:evaluation-image}
\end{equation}
injectivity gives $|S|=n$. The next lemma bounds the character sums on $S$ in terms of the number of omitted rational points.

\begin{lemma}
\label{lem:restricted-character}
Suppose that $h\geq2$. Let $P_1,\ldots,P_n\in X(\Fq)$ be distinct, and let $S\subseteq J$ be their images under the Abel--Jacobi map, as in \eqref{eq:evaluation-image}. Then
\begin{equation}
    \Phi(S)\leq N-n+(2g-2)\sqrt q.
    \label{eq:restricted-character}
\end{equation}
\end{lemma}

\begin{proof}
Fix a nontrivial $\chi\in\widehat J$. Splitting the sum over the curve into the evaluation points and the omitted points gives
\[
\begin{split}
 \sum_{a\in S}\chi(a)
 ={}&\sum_{P\in X(\Fq)}\chi([P-P_0])\\
    &-\sum_{P\in X(\Fq)\setminus\{P_1,\ldots,P_n\}}
                   \chi([P-P_0]).
\end{split}
\]
There are exactly $N-n$ terms in the second sum. Each has absolute value one. The triangle inequality and \eqref{eq:curve-character} therefore give
\[
\begin{split}
 \left|\sum_{a\in S}\chi(a)\right|
 &\leq\left|\sum_{P\in X(\Fq)}\chi([P-P_0])\right|+(N-n)\\
 &\leq(2g-2)\sqrt q+(N-n).
\end{split}
\]
Taking the maximum over all nontrivial characters proves \eqref{eq:restricted-character}.
\end{proof}

Under the dimension assumptions of the main results, $h\geq n\geq2g+2$. Thus Lemma~\ref{lem:restricted-character} applies to every evaluation set used below. It is the only estimate on the evaluation set required in the proof.

\section{Subset Sums and the MDS Condition}
\label{sec:subset-sums}

We first prove an estimate for sums over subsets of a finite abelian group. We then apply it to the set \eqref{eq:evaluation-image}. The use of subsets, rather than tuples with repetitions, is needed because the divisor $B$ in \eqref{eq:mds-vanishing} must be bounded by $D$.

\subsection{A Coefficient Estimate}

For complex numbers $z_1,\ldots,z_n$, set $e_0(z_1,\ldots,z_n)=1$. For $1\leq m\leq n$, define the $m$th elementary symmetric polynomial by
\[
    e_m(z_1,\ldots,z_n)
    =\sum_{1\leq i_1<\cdots<i_m\leq n}z_{i_1}\cdots z_{i_m}.
\]
These polynomials satisfy the identity
\begin{equation}
    \prod_{i=1}^n(1+uz_i)
       =\sum_{m=0}^n e_m(z_1,\ldots,z_n)u^m.
\label{eq:elementary-symmetric}
\end{equation}
We will apply these polynomials to character values. The next bound is an elementary consequence of Cauchy's coefficient formula \cite[Thm.~IV.4]{FlajoletSedgewick2009}. We give a short deduction to specify its constants.

\begin{lemma}
\label{lem:coefficient}
Let $|z_i|=1$ for $1\leq i\leq n$, and let $1\leq m\leq n-1$. Then
\begin{multline}
    \frac{|e_m(z_1,\ldots,z_n)|}{\binom nm}\\
    \leq(n+1)\exp\left\{
      -\frac{m(n-m)}{n^2}
       \left(n-\left|\sum_{i=1}^n z_i\right|\right)
                       \right\}.
\label{eq:coefficient-bound}
\end{multline}
\end{lemma}

\begin{proof}
Put $R=|\sum_{i=1}^n z_i|$ and $r=m/(n-m)$. Apply the arithmetic--geometric mean inequality to the numbers $|1+re^{i\theta}z_i|^2$. Since their mean is at most $1+r^2+2rR/n$, the inequality $1-x\leq e^{-x}$ gives
\begin{multline*}
    \left|\prod_{i=1}^n(1+re^{i\theta}z_i)\right|
       \leq(1+r^2+2rR/n)^{n/2}\\
       \leq(1+r)^n
           \exp\left\{-\frac{r(n-R)}{(1+r)^2}\right\}.
\end{multline*}
Cauchy's coefficient estimate bounds $|e_m|$ by $r^{-m}$ times this maximum modulus. For $p=m/n$, the value $m$ is a mode of the binomial distribution with parameters $n$ and $p$. Thus
\[
 \frac{r^{-m}(1+r)^n}{\binom nm}
 =\frac{1}{\binom nm p^m(1-p)^{n-m}}\leq n+1.
\]
Combining these inequalities and using $r/(1+r)^2=m(n-m)/n^2$ proves the result.
\end{proof}

\subsection{Sums of Distinct Elements}

Let $A$ be a finite abelian group of order $h\geq2$, written additively. Let $S\subseteq A$ have cardinality $n$. For $b\in A$ and $1\leq m\leq n-1$, define
\begin{equation}
    N(m,b,S)=\#\left\{T\subseteq S:
        |T|=m,\ \sum_{a\in T}a=b\right\}.
    \label{eq:subset-count}
\end{equation}
The next proposition gives a sufficient condition for this number to be positive for every $b$. Its form is chosen so that \eqref{eq:restricted-character} can be used directly.

\begin{proposition}
\label{prop:subset-sums}
With the notation above,
\begin{multline}
    \left|\frac{hN(m,b,S)}{\binom nm}-1\right|\\
    \leq(h-1)(n+1)\exp\left\{-\frac{m(n-m)}{n^2}
                         \bigl(n-\Phi(S)\bigr)\right\}.
\label{eq:subset-bound}
\end{multline}
In particular, if
\begin{equation}
    \frac{m(n-m)}{n^2}\bigl(n-\Phi(S)\bigr)
       >\log\bigl((h-1)(n+1)\bigr),
    \label{eq:subset-positivity}
\end{equation}
then $N(m,b,S)>0$ for every $b\in A$.
\end{proposition}

\begin{proof}
Write $S=\{a_1,\ldots,a_n\}$. Character orthogonality gives
\begin{equation}
    hN(m,b,S)=
    \sum_{\chi\in\widehat A}\chi(-b)
       e_m\bigl(\chi(a_1),\ldots,\chi(a_n)\bigr).
    \label{eq:Fourier-inversion}
\end{equation}
The trivial character contributes $\binom nm$. For each nontrivial character, Lemma~\ref{lem:coefficient} and the definition of $\Phi(S)$ give
\begin{multline}
    \left|e_m\bigl(\chi(a_1),\ldots,\chi(a_n)\bigr)\right|\\
    \leq\binom nm(n+1)
       \exp\left\{-\frac{m(n-m)}{n^2}
                         \bigl(n-\Phi(S)\bigr)\right\}.
\label{eq:nontrivial-coefficient}
\end{multline}
There are $h-1$ such characters. The triangle inequality proves \eqref{eq:subset-bound}. Under \eqref{eq:subset-positivity}, its right-hand side is less than one. Thus $hN(m,b,S)/\binom nm>0$ for every $b$.
\end{proof}

\subsection{Application to the MDS Condition}

We return to $X$, $D$, and $G$. The preceding proposition counts divisors supported on distinct evaluation points after applying the Abel--Jacobi map. To contradict the MDS condition, it is enough to find one such divisor $B$ with $G-B$ equivalent to a fixed effective divisor of degree $g-1$. The next proposition makes this step precise and retains the actual numbers of rational points of the curve and its Jacobian.

\begin{proposition}
\label{prop:necessary}
Suppose that $C_{\mathcal L}(D,G)$ is MDS and that $k,n-k\geq g+1$. With $N$ and $h$ as in \eqref{eq:J-h-N}, one has
\begin{equation}
\begin{split}
    \frac{k(n-k)}{n^2}
       \bigl(2n-N-(2g-2)\sqrt q\bigr)
    \leq\log\bigl((h-1)(n+1)\bigr).
\end{split}
\label{eq:intrinsic-necessary}
\end{equation}
Consequently,
\begin{equation}
\begin{split}
    \frac{k(n-k)}{n^2}
       \bigl(2n-q-1-(4g-2)\sqrt q\bigr)
    \leq\Lambda_g(q).
\end{split}
\label{eq:uniform-necessary}
\end{equation}
\end{proposition}

\begin{proof}
Proposition~\ref{prop:code-divisors} gives $\deg G=k+g-1$. Take $P_0$ as in \eqref{eq:Abel-Jacobi}, let $S$ be the set in \eqref{eq:evaluation-image}, and put
\begin{equation}
    b=[G-(k+g-1)P_0]\in J.
    \label{eq:target-class}
\end{equation}
The points in $S$ are distinct. In particular, $h\geq n\geq2g+2$, so the logarithm in \eqref{eq:intrinsic-necessary} is defined.

Suppose that $N(k,b,S)>0$. The corresponding subset of evaluation points gives a divisor $0\leq B\leq D$ of degree $k$ such that
\[
    [B-kP_0]=[G-(k+g-1)P_0].
\]
It follows that
\begin{equation}
    G-B\sim(g-1)P_0.
    \label{eq:effective-residual}
\end{equation}
The divisor $(g-1)P_0$ is effective for every $g\geq1$; for $g=1$, it is the zero divisor. Thus $\ell(G-B)\geq1$, contrary to \eqref{eq:mds-vanishing}. We have proved
\begin{equation}
    N(k,b,S)=0.
    \label{eq:missing-subset}
\end{equation}

Apply \eqref{eq:subset-bound} with $A=J$ and $m=k$. By \eqref{eq:missing-subset}, its left-hand side equals one. Taking logarithms yields
\begin{equation}
    \frac{k(n-k)}{n^2}\bigl(n-\Phi(S)\bigr)
       \leq\log\bigl((h-1)(n+1)\bigr).
    \label{eq:necessary-Phi}
\end{equation}
By Lemma~\ref{lem:restricted-character},
\[
    n-\Phi(S)\geq2n-N-(2g-2)\sqrt q,
\]
we obtain \eqref{eq:intrinsic-necessary}. Finally, \eqref{eq:Weil} and $n\leq N$ give
\begin{align*}
    N+(2g-2)\sqrt q&\leq q+1+(4g-2)\sqrt q,\\
    \log\bigl((h-1)(n+1)\bigr)&\leq\Lambda_g(q).
\end{align*}
These inequalities imply \eqref{eq:uniform-necessary}.
\end{proof}

\section{Proofs of the Main Results}
\label{sec:proof}

Proposition~\ref{prop:necessary} is stronger than the uniform bound when the curve and its Jacobian have fewer rational points than their Weil upper bounds. We now use the uniform form to prove Theorem~\ref{thm:main} and Corollaries~\ref{cor:epsilon}--\ref{cor:fixed-genus}. The sign of the expression involving $n$ is separated before the factor $k(n-k)$ is estimated.

\begin{proof}[Proof of Theorem~\ref{thm:main}]
For this proof, write
\[
    T=q+1+(4g-2)\sqrt q,\qquad \Lambda=\Lambda_g(q).
\]
If $n\leq T/2$, then \eqref{eq:main-bound} follows because $0<1-\Lambda/\kappa<1$. Suppose that $n>T/2$. Since $\kappa\leq n/2$,
\begin{equation}
    k(n-k)=\kappa(n-\kappa)\geq\frac{\kappa n}{2}.
    \label{eq:dimension-product}
\end{equation}
The factor $2n-T$ is positive. Therefore \eqref{eq:uniform-necessary} and \eqref{eq:dimension-product} imply
\begin{equation}
    \kappa\left(1-\frac{T}{2n}\right)\leq\Lambda.
    \label{eq:rearrangement}
\end{equation}
As $\kappa>\Lambda$, rearranging \eqref{eq:rearrangement} gives
\[
    n\leq\frac{T}{2(1-\Lambda/\kappa)},
\]
which is \eqref{eq:main-bound}.
\end{proof}

It remains to prove the three corollaries. Corollary~\ref{cor:epsilon} follows from the denominator in the main theorem. Corollary~\ref{cor:rate} uses \eqref{eq:rearrangement} before division by $1-\Lambda/\kappa$. Corollary~\ref{cor:fixed-genus} then follows by estimating $\Lambda_g(q)$ for fixed $g$.

\begin{proof}[Proof of Corollary~\ref{cor:epsilon}]
The assumption \eqref{eq:epsilon-range} gives $\kappa\geq g+1$ and
\[
    \frac{\Lambda_g(q)}{\kappa}
       \leq\frac{2\varepsilon}{1+2\varepsilon}<1.
\]
Hence Theorem~\ref{thm:main} applies, and
\[
    \frac{1}{2(1-\Lambda_g(q)/\kappa)}
       \leq\frac{1+2\varepsilon}{2}.
\]
This proves \eqref{eq:epsilon-bound}. For fixed $g$,
\[
    \Lambda_g(q)=(g+1)\log q+O_g(q^{-1/2}).
\]
The ceiling in \eqref{eq:epsilon-threshold} changes its value by less than one, and the maximum with $g+1$ is immaterial for sufficiently large $q$. This proves \eqref{eq:threshold-asymptotic}.
\end{proof}

\begin{proof}[Proof of Corollary~\ref{cor:rate}]
Again put $T=q+1+(4g-2)\sqrt q$ and $\Lambda=\Lambda_g(q)$. If $n\leq T/2$, the conclusion is immediate. If $n>T/2$, the argument leading to \eqref{eq:rearrangement} uses only $\kappa\geq g+1$, not $\kappa>\Lambda$. Thus
\[
    n\leq\frac T2+\frac{n\Lambda}{\kappa}
      \leq\frac T2+\frac{\Lambda}{\eta}.
\]
This proves \eqref{eq:rate-bound}.
\end{proof}

\begin{proof}[Proof of Corollary~\ref{cor:fixed-genus}]
For fixed $g$, definition \eqref{eq:Lambda} gives
\[
    \Lambda_g(q)=(g+1)\log q+O_g(q^{-1/2})
                =O_g(\log q).
\]
Since $\eta>0$ is fixed,
\[
    \frac12+\frac{\Lambda_g(q)}{\eta}
       =O_{g,\eta}(\log q).
\]
Substituting this estimate into \eqref{eq:rate-bound} proves \eqref{eq:rate-asymptotic}.
\end{proof}

\begin{remark}
\label{rem:scope}
For fixed $g$, Theorem~\ref{thm:main} implies
$\limsup n/q\leq1/2$ along any sequence of MDS codes for which $q\to\infty$ and $\kappa/\log q\to\infty$. This statement is an upper bound. No matching construction for every fixed genus is proved here. Genus zero is excluded throughout: generalized Reed--Solomon codes already have length $q$ and extended versions have length $q+1$.
\end{remark}

\section{Future Work}
\label{sec:future}

The first question is whether the dependence on $\log q$ in the dimension condition can be reduced. The present proof estimates every nontrivial character separately and then sums the absolute values over the Jacobian. This introduces a factor of order $q^g$. The exponential term in \eqref{eq:subset-bound} must dominate that factor. It would be useful to obtain a bound near $q/2$ under only $g+1\leq k\leq n-g-1$, or to identify curves and divisors that prevent such a statement. The proof in this paper does not decide this question.

A second question concerns the term of order $\sqrt q$. In Corollary~\ref{cor:rate}, its coefficient is $2g-1$. This coefficient comes from bounding the number of rational points and the nontrivial character sums separately. Proposition~\ref{prop:necessary} retains the actual values of $|X(\Fq)|$ and $|\Jac(X)(\Fq)|$. Further information about the curve or its character sums may improve this bound. Such an improvement would be relevant even if the dimension condition is unchanged.

A third question is whether the coefficient $1/2$ is optimal for each fixed genus $g\geq2$ in the dimension ranges considered here. This requires constructions of MDS codes whose lengths approach $q/2$, or an upper bound with a smaller coefficient. It is also natural to study the dependence on $g$ when the genus grows with $q$. The explicit form of Theorem~\ref{thm:main} gives a starting point, but it does not establish sharpness in any such regime.

\section*{Acknowledgment}

ChatGPT 6 (OpenAI) was used for literature searches, mathematical discussions, and the polishing of this manuscript.

The work of Haojie Chen was supported by the National Natural Science Foundation of China under Grant 12441107, the Guangdong Basic and Applied Basic Research Foundation under Grant 2025A1515011764, and the National Key Research and Development Program of China under Grant 2025YFA1017100. The work of Chuangqiang Hu was supported by the National Natural Science Foundation of China under Grant 12441107.

\end{document}